# Ocean Worlds Exploration
## and the
## Search for Life

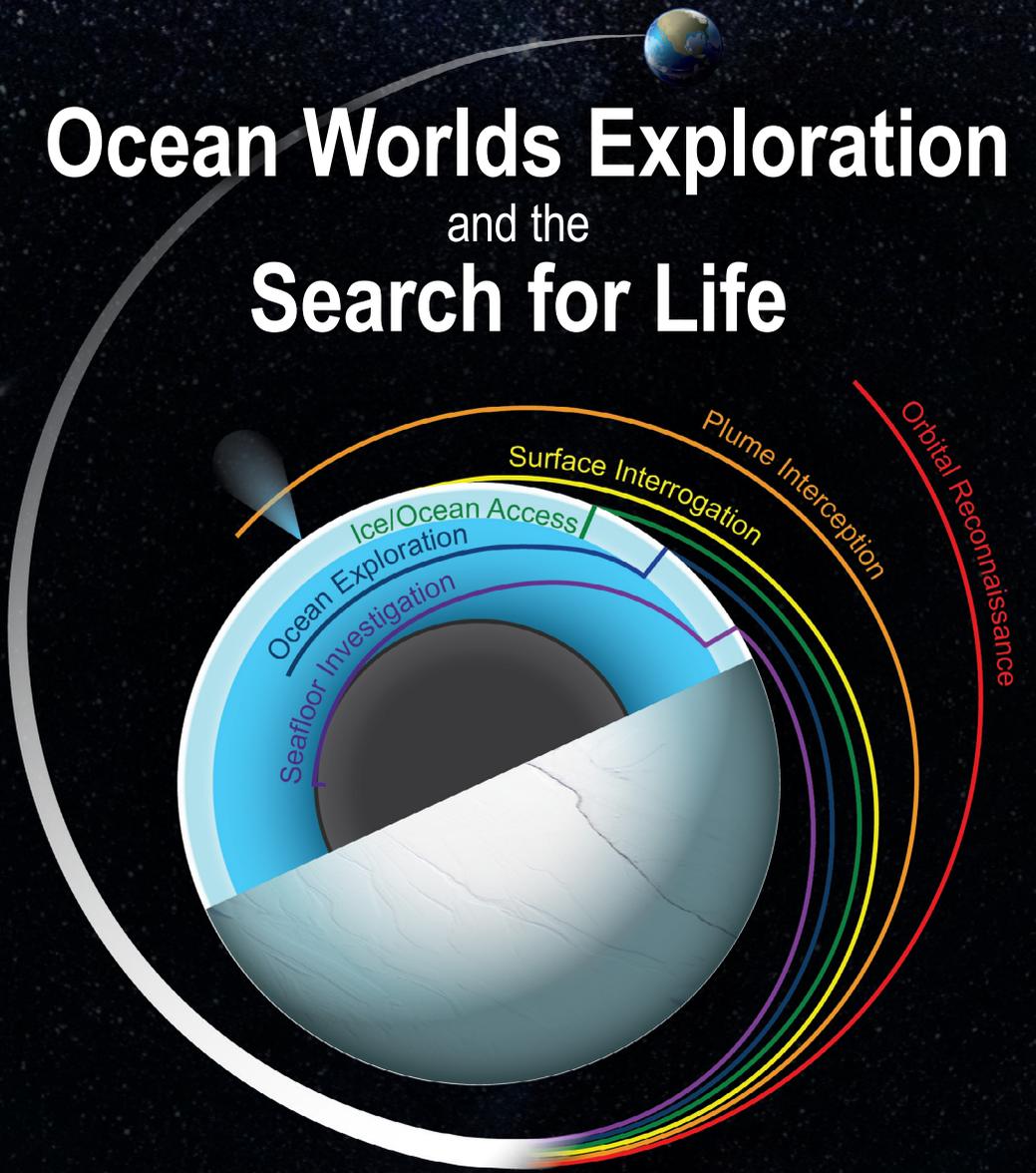

A White Paper reflecting the views of NASA's Network for Ocean Worlds,
submitted to the Decadal Survey in Planetary Science and Astrobiology

LEAD AUTHORS

**Samuel M. Howell*** 
NASA Jet Propulsion Laboratory
California Institute of Technology

**William C. Stone**
Stone Aerospace

**Kate Craft**
Johns Hopkins University
Applied Physics Laboratory

*Contact: samuel.m.howell@jpl.nasa.gov | 818.393.8213

NOW CO-LEADS

**Christopher German**
Woods Hole Oceanographic
Institution

**Alison Murray**
Desert Research
Institute

**Alyssa Rhoden**
Southwest Research
Institute

**Kevin Arrigo**
Stanford
University

A list of the > 180 individuals who have endorsed this white paper can be found at: **oceanworlds.space/whitepaper**

## SUMMARY

We recommend the establishment of a dedicated **Ocean Worlds Exploration Program** within NASA to provide sustained funding support for the science, engineering, research, development, and mission planning needed to implement a multi-decadal, multi-mission program to explore Ocean Worlds for life and understand the conditions for habitability. The two new critical flagship missions within this program would 1) land on Europa or Enceladus in the decade 2023-2032 to investigate geophysical and geochemical environments while searching for biosignatures, and 2) access a planetary ocean to directly search for life in the decade 2033-2042. The technological solutions for a landed mission are already in-hand, evidenced by the successful delta-Mission Concept Review of the Europa Lander pre-flight project in the fall of 2018. Following an initial landed mission, an ocean access mission will require substantial research, development, and analog testing *this* decade to enable the initiation of a pre-flight project at the start of the *following* decade.

This ambitious goal could not have been put forward with comparable, requisite credibility in any preceding decade. **Both science and technology have now matured so that we may prioritize the direct search for signs of life**, applying lessons learned from planetary and Earth exploration to *in situ* investigations of worlds where water and life are most likely to exist today.

This new era of planetary investigation will require significant support for cross-disciplinary research and development, bringing together planetary scientists and engineers with those that study the Earth. To chart these new waters, a diversity of disciplines must be accompanied by a diversity of experience, skill, and perspective. All facets of this Ocean Worlds Exploration Program must actively ensure equitable access and inclusion across the spectrum of contributors in Planetary Science, Exploration, and Astrobiology. Hence, we recommend an Ocean Worlds Exploration Program that includes specific initiatives designed to entrain and retain scientists and technologists from historically under-represented groups and to support a healthy work-life balance that **enables equitable participation from across the entire community.**

An Ocean Worlds Exploration Program will provide opportunities for advancing exciting technologies to climb "the ladder for life detection," a framework to further the search for life that begins with a foundation in habitability (*1*). This program will spur a new era of innovation within NASA as it pursues one of the most fundamental science quests of the Science Mission Directorate (*2*) and humanity as a whole: the search for life elsewhere. We champion the value of Ocean Worlds exploration as core to NASA's Science Strategy, and as the best path to "*Challenge assumptions about what is technically feasible and enable revolutionary scientific discovery through a deliberate focus on innovation, experimentation, and cross-disciplinary research* (*2*)."

## SCIENCE DRIVERS

As a NASA Astrobiology Program Research Coordination Network (RCN), The *Network for Ocean Worlds* (NOW) was established to advance comparative studies to characterize Earth and other Ocean Worlds across their interiors, oceans, and cryospheres; to investigate their habitability; to search for biosignatures; and to understand life—in relevant ocean world analogs and beyond.

The recognition that life can be chemically sustained, and possibly first emerged, in the darkest reaches of Earth's deep ocean (*3*) made the discovery of a deep global saltwater ocean beneath Europa's icy shell a watershed moment in Planetary Science and Astrobiology (*4*). Two decades later, the Ocean Worlds of the outer solar system now present the best opportunity to find life in several possible states: metabolically alive and active, alive but dormant, or life-like in form but no longer living (*1*). There is also the possibility that Ocean Worlds possess pre-biotic environments where life has not yet emerged, providing a pristine look into early crucibles for life.

**The excitement associated with Ocean Worlds is attracting a diverse new cohort of scientists** who have not traditionally been involved in solar system exploration: experts in Earth's oceanography, glaciology, polar environments, and extreme-environment microbiology. Planetary scientists from established disciplines such as geophysics, geochemistry, orbital dynamics, and astronomy are now working closely with terrestrial scientists to meaningfully advance theoretical modelling, comparative planetology, laboratory experiments, and analog field studies. Detecting life on ocean worlds is also synergistic with the Network for Life Detection (NfoLD) RCN and studies of Ocean Worlds offer a compelling opportunity to challenge our understanding of the habitable regions of our Solar System (including Earth) and exoplanetary systems, relevant to the goals of the Nexus for Exoplanet System Science (NExSS) RCN.

The oceans of Europa, Enceladus, and Titan have emerged as key targets in the search for extant life (as opposed to past life) in the Solar System (*5*). The icy shell of Titan is likely too thick and hazardous for near-term robotic exploration of the ocean, where habitable environments are most likely (*6*), but the icy outer shells of Europa and Enceladus, which may be a few kilometers to a few tens-of-kilometers thick, provide surmountable new environments to explore (*7–12*).

Europa and Enceladus possess global oceans in direct contact with rocky interiors, providing the potential for water-rock reactions and a range of environments in which life could emerge and persist (*1*). Importantly, they also exhibit ongoing geologic activity, driven in part by gravitational interactions with their respective host planets, permitting surface-interior material transport that can maintain chemical disequilibria in the ocean and at its interfaces. **Thus, the key components for life are present: water, chemical building blocks in disequilibrium, energy, and time**.

*Table 1 | NOW priority themes and questions in Planetary Science and Astrobiology.*

| | |
|---|---|
| Life and Habitability | Has life emerged within Ocean Worlds, and does it persist today? |
| | What are the habitable environments and interfaces of Ocean Worlds? |
| | How do Ocean Worlds inform our assessments of the habitability of exoplanet systems? |
| Planetary Formation | How did Ocean Worlds form, and how did they accrete their volatiles? |
| | How are organics delivered to and synthesized on Ocean Worlds, and how do they evolve? |
| | What can the chemical building blocks of Ocean Worlds tell us about the evolution and transport history of these materials across the Solar System? |
| Planetary Evolution | What are the key interfaces that permit and regulate thermal, physical, and chemical exchange? |
| | How did differentiation occur for the key geologic layers: the ice shell, ocean, and rocky interior? |
| | What tidal interactions and other sources of energy power change within these layers? |
| Comparative Oceanography | What analog geological processes on Earth might provide insight into the sources of chemical energy that may render an Ocean World habitable? |
| | How do the diversity of chemical reactions that arise on Ocean Worlds fuel potential life-supporting metabolisms, and how do they vary in time and space? |
| | How and on what timescales are biosignatures transported, modified, and preserved? |

Embodied in the search for life is a focus on the planetary and local conditions that enable (or restrict) the emergence and persistence of life. This includes exploring the range of habitable environments present in the outer Solar System and expanding the techniques for accessing potential habitats and identifying signs of past or present life. Hence, the search for life among Ocean Worlds can fulfill a major role in both Planetary Science and Astrobiology if it simultaneously addresses key questions related to planetary system and body processes, focusing on the physical and chemical processes that shape these environments (**Table 1**).

**PRIORITY MISSIONS FOR THE DECADE 2023-2032**

Here we outline a systematic pathway to ocean access and life detection by 2042 that will build upon current flight projects to explore Ocean Worlds (**Figure 1**). These include the flagship *Europa Clipper* fly-by mission to investigate the habitability of Europa, and the New Frontiers *Dragonfly* mission to traverse the surface of Titan, a first in the outer solar system. Contemporaneously, ESA's L-class *JUICE* mission will be the first designed to assess the relative habitability of multiple Ocean Worlds within a single planetary system. Timely completion and launch of *Europa Clipper* and *Dragonfly*, and sustained international partnership in relation to the *JUICE* mission, are central to advancing Planetary Science and Astrobiology in the next decade.

**The most critical new mission in the coming decade required to progress towards life detection is a landed mission to either Europa or Enceladus**. The choice between first visiting Europa or Enceladus for surface and interior exploration should be made according to knowledge of each target at the time of selection, including habitability and potential for extant life, and technology and mission readiness. The Europa Lander Science Definition Team report (*14*) and Enceladus Orbilander Flagship Planetary Mission Concept Study (*15*), among others, provide strong science rationales for biosignature detection and environmental investigation of each planetary surface. Chemistry resulting from life may be retained in the non-ice materials of an Ocean World's icy shell (e.g. organics structurally modified by life, by-products of metabolism), providing compelling evidence for the existence of life, even in the absence of lifeform detection. Detecting these biosignatures is a priority science objective for any *in situ* Ocean Worlds mission.

In addition to the search for biosignatures on an Ocean World, a landed mission is crucial for investigating geophysical and geologic context and ground-truthing remote sensing data. Well-defined measurements will provide a strong foundation for inferences regarding the balance of physical and chemical processes in the ice and ocean, as well as their influences on habitability (*14*, *15*). Even in the case of a non-biosignature detection, surface measurements will continue to be valuable for considering the influence of landing site selection, radiation processing, and geologic and geochemical context on biosignature prevalence and degradation.

A landed mission on Enceladus or Europa will also be enabling for future ocean access and direct life detection in a planetary ocean, where the most habitable interfaces are likely located (*6*). Characterizing the environments on the surfaces of and within Ocean World ice shells will provide key engineering and design constraints for ice and ocean access technologies under development for deployment in the 2030s, and retire the risks associated with landing and deploying such payloads. In 2018, the Europa Lander pre-flight project passed its delta Mission Concept Review, proving Pre-Phase A readiness to land on an icy airless body with unknown lander-scale surface topography in the outer solar system. **We therefore advocate that a landed flagship mission capable of performing *in situ* biosignature detection and ascertaining the geophysical and geochemical conditions on Europa or Enceladus advance to Phase A as rapidly as possible.**

After a landed flagship mission, the next highest priority Ocean Worlds missions for the new decade are an Enceladus or Europa plume fly-through mission, and/or a Titan global surveyor mission. Recent proposals have already demonstrated that these concepts can be accomplished under the New Frontiers cost cap. **Therefore, we strongly recommend that Ocean Worlds, including Titan, be retained as permitted targets under the New Frontiers program.**

*Plume Flythrough:* The so-far unparalleled cryovolcanic activity at the south pole of Enceladus (*16*) and growing case for water vapor plumes at Europa (*17, 18*) offer windows into the habitability, chemistry, and geologic processes within these bodies. Sampling material from these plumes and characterizing any corresponding mechanical or geodetic changes in the ice shell (*19*) will allow high-fidelity investigations of interior volatile reservoirs and their habitability, illuminate geologic environments and processes, and potentially allow for biosignature detection.

*Titan Global Surveyor:* Titan is key to understanding how distributed organic materials and an active methane cycle react with water to potentially create chemical nutrients and building blocks for life. The selected New Frontiers *Dragonfly* mission will make critical point measurements at several locations near an impact crater and dune fields on Titan. Observations by a future orbiter or atmospheric mission would place the *Dragonfly* point measurements in larger context with respect to Titan's geologically diverse terrains and provide insights to the co-evolution of the atmosphere and surface. This will illuminate the coupling between the surface organics and the vast subsurface ocean—a potentially organic-rich habitable environment.

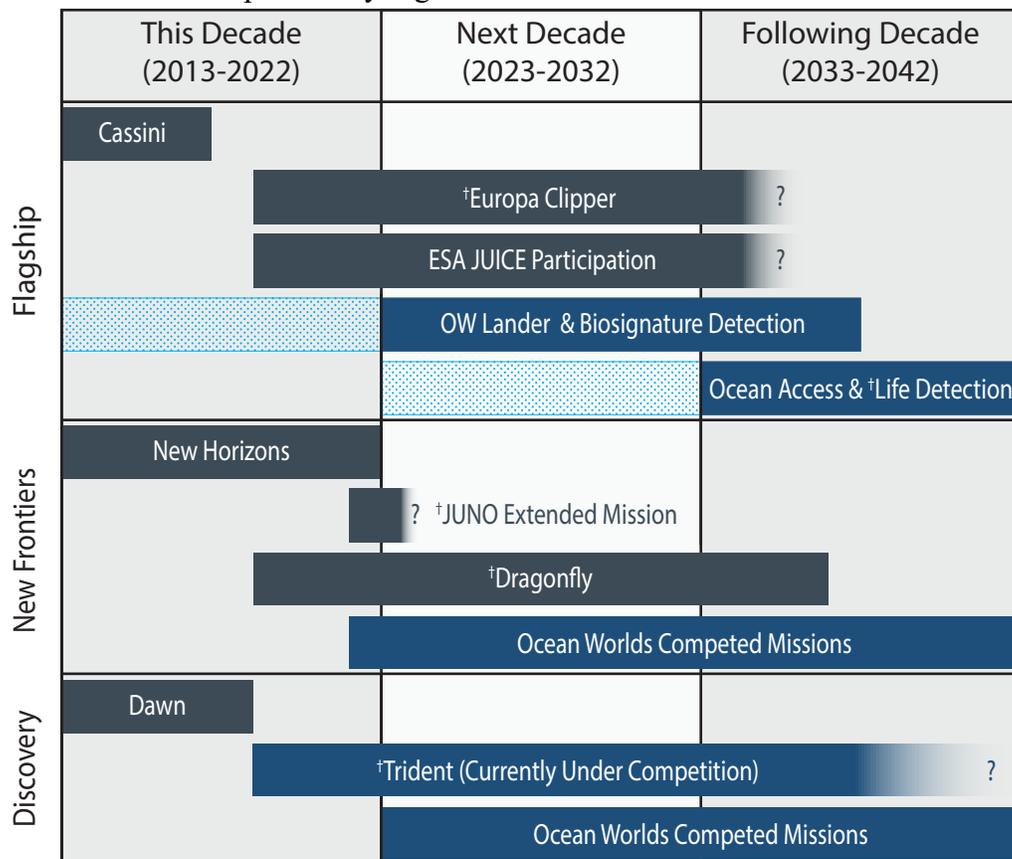

*Figure 1 | The NOW-recommended Ocean Worlds Exploration Program leading to Ocean Access and Life Detection in the Decade 2033-2042. Grey boxes depict funded/flown missions, solid blue boxes depict new missions, and blue hashed boxes represent research, analysis, and technology maturation.*
† *These NOW priorities have also been identified by OPAG (13).*

While the priority for an Ocean Worlds Exploration Program should be to pursue *in situ* studies on confirmed Ocean Worlds to advance the search for life, **the program should also seek to verify and explore potential Ocean Worlds elsewhere in the Solar System**. Foremost is Neptune's moon Triton (*20*), which exhibits a geologically young surface, dynamic geologic environment, and potential for sustained interior activity. The moons of the ice giant systems, which may harbor oceans, could be investigated within the cost cap of a Discovery-class mission, as demonstrated by the proposed Trident mission (*20*), or as part of a larger ice giants system mission in the New Frontiers mission class (*21*) or as a Flagship (*13*). In the case of a broader ice giants system mission, the primary science objectives should include a significant focus on ocean detection and characterization for comparison to previously discovered oceans (Europa, Enceladus, Titan, Ganymede, Callisto) in order to advance our understanding of physical and chemical processes that control the relative habitability among the diverse set of Ocean Worlds.

Despite the scientific potential of Ocean Worlds, programmatic constraints are insulating the exploration of these targets from innovations in spacecraft technologies. Low cost, tailored investigations that address questions cited in **Table 1** would increase the cadence of Ocean Worlds science return, allowing future investigations to respond to new discoveries. We therefore recommend that considerations be made to permit the exploration of outer solar system targets in lower-cost-cap, Small Satellite (SmallSat) mission classes (e.g. SIMPLEx).

We recommend removing Phase E costs from the cost caps associated with these programs to make cost comparisons more agnostic to mission cruise requirements. Specifically, the long-duration of outer solar system missions requires long inter-planetary transits, increasing the programmatic costs. Additionally, we recommend that NASA establish an aggressive life-testing program for SmallSat technologies to extend reliability to outer solar system mission durations, as well as a program for SmallSat planetary protection considerations. These analyses are cost prohibitive on a case-by-case basis. SmallSats may be carried to their destination by a primary host spacecraft to improve science return. Reserving resources within larger mission calls for opportunistic SmallSat "ride-alongs" may greatly increase the accessibility of Ocean Worlds.

## RESEARCH, ANALYSIS, AND TECHNOLOGY DEVELOPMENT

**We recommend that NASA establishes a sustained research and development program in the coming decade (2023 – 2032) for scientific research, deep subsurface ice and ocean exploration technologies, and instrumentation, within a broader Ocean Worlds Exploration Program, to enable ocean access and life detection in the following decade (2032 – 2042).**

*Research and Analysis*

At the core of Ocean Worlds exploration in the coming decades should be a well-funded and sustained research and analysis (R&A) program. Priorities should include continued support for analyses of past, current, and future mission data, while encouraging new and novel studies into Ocean World processes—including the emergence, persistence, and biogeochemical ramifications of life. This approach will broaden our understanding of the key measurements needed to confirm life beyond Earth. Supporting theoretical, laboratory, modeling, and terrestrial investigations into comparative oceanography, ice physics, and multi-disciplinary cryosphere research will build an empowered community capable of answering key questions to guide future exploration.

We are fortunate to live on and explore an Ocean World. Strong synergies between icy moon exploration and the studies of polar oceans and ices on Earth can expand our capabilities for interrogating these environments throughout the Solar System. An Ocean Worlds Exploration Program geared for success in the search for life must devote significant efforts to developing exploration technologies and instruments (and associated detection limits), which will include a strong analog test component (*11*). We recommend that analog research and field testing of exploration technologies and instrumentation continue under expanded and sustained funding.

As we understand the increasing likelihood that Ocean Worlds are common in other solar systems, accounting for potentially 1-in-4 volcanically active exoplanets (*22*), fostering collaborations with exoplanet researchers may yield new and challenging ideas about the habitability of other worlds. In addition to studying extrasolar systems, ground-based (*19*) and space-based (*18*) telescopes can augment Ocean Worlds missions in order to study time-series phenomena, provide global context, and investigate the properties of yet-unexplored surfaces.

*Exploration Technology Development*

In 2017, NASA commissioned a Keck Institute for Space Studies (KISS) investigation into subsurface access on Ocean Worlds (*7*). This study surveyed more than half-a-century of research into deep-ice access technologies (*12*) in the context of our current understanding of the icy environments of the outer Solar System. One technology concept emerged to be robust against the potential hazards and challenges of planetary exploration: the ice penetrating robot, or "**cryobot**," that can effectively penetrate ice shells tens-of-kilometers thick, starting initially in vacuum at cryogenic temperatures and descending under control through the ice column, into and through an ocean water column. **The 2017 KISS study concluded that for the first time, the science and technology of ocean access have reached a level of maturity to credibly enable flight implementation within the next two decades**. NASA's Science Mission Directorate then took the next step towards ocean access by establishing the Scientific Exploration Subsurface Access Mechanism for Europa (SESAME) program. Cryobot technologies to enable planetary ocean access missions are now in the engineering development phase (*8–11*).

Ocean access technologies require robust development, integration, and analog testing this coming decade to enable flight readiness in the 2030s (*7–11*). Cryobot development priorities include heat and power systems, through-ice and in-ocean communications, thermal management, autonomy, mobility, scientific instrumentation, and planetary protection. In addition, under-ice and in-ocean mobility technologies (e.g. cryobot mobility or cryobot-deployed payloads) are vital to amplify the probability of life detection in icy and ocean environments.

*Instrument Development*

Sustained development is needed for remote, *in situ,* and aqueous instrument concepts to ensure that biosignature detection and habitability characterization devices honed for laboratory and terrestrial field analyses can be adapted to use in space, in harsh planetary environments (e.g. high radiation at Europa), and with samples that are water-based rather than soil or dust-derived. Several NASA technology programs (e.g. ICEE, ICEE2, COLDTech, PICASSO, MatISSE) have *initiated* support for technology development with respect to Ocean Worlds-relevant science, but only for a few life-detection instrument types thus far. There must be a concerted effort to both introduce new technologies and further the TRL of life detection instruments. This includes instruments for detecting physical signs of life in addition to chemical attributes of non-ice materials that could contain life or life-derived byproducts. Additional efforts are required to mature capabilities for sample preparation, collection, and transport (e.g. microfluidics), in order for planetary missions seeking signs of life to be successful. To this end, NASA should accelerate high risk, high reward investment in development of novel instruments, sampling, and sample access systems.

*Thermal and Electrical Power Systems*

Accessible liquid water oceans beyond Earth are exclusive to the outer Solar System. Therefore, ensuring the steady production of $^{238}$Pu for use in spacecraft power systems is critical to enabling the search for life (*10*, *13*, *23*). Additionally, subsurface exploration and ocean access technologies require the availability of high-heat density radioisotope power systems, which may require the repackaging of TRL-9 RTG components (*8–10*, *23*), or low-volume, low-mass fission power systems (*8*). Therefore, NASA must work with the Department of Energy to secure the availability of high heat density thermal and electrical power systems for flight by 2033.

## STRATEGIC PARTNERSHIPS

Coordinated Ocean Worlds studies at NASA will connect stakeholders across the Planetary Science and Earth Science Divisions of the Science Mission Directorate, spurring interdisciplinary research. Internationally, outer solar system missions are exclusive to the mature space agencies NASA and ESA. However, ISRO and JAXA now have the science, technology, and mission planning readiness to execute Ocean Worlds missions, having implemented exciting, low-cost inner solar system missions. NASA should encourage international partners to further Ocean Worlds exploration by sustaining funded programs for domestic scientists to collaborate, leveraging international funding and ideas that will benefit the community as a whole.

Terrestrially, NSF's Geosciences Directorate provides the main source of funded access to the Earth's cryosphere—a key zone for validation of Ocean Worlds exploration technologies and scientific payloads. Additionally, NOAA coordinates a multiple-agency, multiple-partner national program of Ocean Exploration and Research that offers opportunities to access unexplored regions of Earth's oceans, while the International Ocean Discovery Program emphasizes deep ocean life, creating the potential for leveraged funding. Further, the National Oceanographic Partnerships Program provides an example of an inter-agency mechanism to facilitate industry collaborations on oceanic research, with signatories that include NASA, NSF, and NOAA.

**We recommend that NASA, NSF, and NOAA coordinate a national program for Ocean Worlds studies** that capitalizes on existing investments across agencies, incentivizing government and the private sector to embrace their diverse expertise in exploration, life detection, and comparative oceanography. This requires sustained, robust access to the Antarctic ice sheet and its subglacial water bodies for the purposes of testing access technologies and science payloads.

**(1)** *Neveu+ (2018) doi:10.1089/ast.2017.1773*; **(2)** *NASA (2020) link*; **(3)** *Martin+ (2008) doi: 10.1038/nrmicro1991*; **(4)** *Khurana+ (1998) doi:10.1038/27394*; **(5)** *Sherwood+ (2018) doi:10.1016/j.actaastro.2017.11.047*; **(6)** *NAS (2018) doi:10.17226/25252*; **(7)** *Sotin+ (2017) link*; **(8)** *†Schmidt+ (2020) Subsurface Needs for Ocean Worlds*; **(9)** *†Fleurial+ (2020) Planetary Ocean Access: Technology Needs and Priorities for Closing a Cryobot System Architecture*; **(10)** *†Woerner+ (2020) Radioisotope Heat Sources and Power Systems for Enabling OW Subsurface and Ocean Access Missions*; **(11)** *†Stone+ (2020) National Ocean Worlds Analog Field Observatory*; **(12)** *Stone+ (2017) doi: 10.1007/978-3-319-73845-1_4*; **(13)** *†Moore+ (2020) ArXiv200311182*; **(14)** *Hand+ (2017) link*; **(15)** *†MacKenzie+ (2020) The Enceladus Orbilander: A Flagship-class Mission Concept for Astrobiology*; **(16)** *†Ermakov+ (2020) A recipe for geophysical exploration of Enceladus*; **(17)** *Schenk+ (2018) Enceladus ISBN:978-0816537075*; **(18)** *Jia+ (2018) doi: 10.1038/s41550-018-0450-z*; **(19)** *Paganini+ (2020) doi:10.1038/s41550-019-0933-6*; **(20)** *Prockter+ (2018) EPSC*; **(21)** *†Leonard+ (2020) A New Frontiers Class Mission for the Uranian System that Focuses on Moon, Magnetosphere, and Ring Science*; **(22)** *Quick+ (2020) doi:10.1088/1538-3873/ab9504*; **(23)** *†Balint+ (2020) Enabling the Next Frontiers in Astrobiology—Ocean and Ice Worlds Exploration with a Radioisotope Power System Inside a Pressure Vessel* | *†***Denotes Decadal White Paper**